\documentclass[twocolumn,amsmath,amssymb,prl,showpacs]{revtex4}

\usepackage[dvips]{color}

\newcommand{\nc}{\newcommand}

\nc{\tcb}{\textcolor{blue}}
\nc{\tcr}{\textcolor{red}}

\nc{\ov}{\overline}

\def\be{\begin{equation}}
\def\ee{\end{equation}}
\def\bea{\begin{eqnarray}}
\def\eea{\end{eqnarray}}

\begin{document}

\title{CPT-violating neutrino oscillations}

\author{S. Esposito}
\affiliation{Dipartimento di Scienze Fisiche, Universit\`a di Napoli ``Federico II'' and \\ Istituto Nazionale di Fisica Nucleare, Sezione di Napoli \\
Complesso Universitario di Monte S. Angelo,  Via Cinthia, I-80126
Naples, Italy}

\author{G. Salesi}
\affiliation{Universit\`a di Bergamo, Facolt\`a di
Ingegneria, viale Marconi 5, 24044 Dalmine (BG), Italy
\\ and Istituto Nazionale di Fisica Nucleare, Sezione di Milano
via Celoria 16, I-20133 Milan, Italy}

\begin{abstract}
\noindent We propose a simple phenomenological model predicting, through Lorentz symmetry breaking, a CPT-violating asymmetry between particle and antiparticle states in neutrino oscillations involving sterile neutrinos. Such a model is able to explain the apparently observed anomalous excess of low-energy $\nu_e$-like events, reported by the MiniBooNE collaboration, as well as the non-observation of the corresponding anomalous excess of $\ov{\nu}_e$-like events.
The present model leads to very specific physical predictions in the neutrino oscillations scenario, and account for the observed anomalies in terms of only one CPT-Lorentz violation parameter of the order of the Grand-Unification energy scale.
\end{abstract}

%11.30.Cp Lorentz and Poincar\'e invariance
%11.55.Fv Dispersion relations
%13.15.+g Neutrino interactions
%14.60.St Non-Standard-Model neutrinos, right-handed neutrinos, etc.
%14.60.Pq Neutrino mass and mixing (see also 12.15.Ff Quark and lepton masses and mixing)
%26.65.+t Solar neutrinos (see also 96.60.Vg Particle emission, solar wind in solar physics)
\maketitle

%%%%%%%%%%%%%%%%%%%%%%%%%%%%%%%%%%%%%%%%%%%%%%%%%%%%%%%%%%%%%%%%%%%%%%%%%%%%%%%%%%%%%%%%%%%%%%%%%%%%%

\noindent In recent years, compelling evidence for the phenomenon of neutrino oscillations \cite{NuOsc} has been firmly established by atmospheric \cite{BE12}, solar \cite{BE345}, reactor \cite{BE678} and long-baseline accelerator \cite{BE910} neutrino experiments. With a confidence level of more than $7 \sigma$, indeed, we now know that solar $\nu_e$'s convert to $\nu_\mu$ or $\nu_\tau$, while atmospheric $\nu_\mu$'s disappear (most likely converting to $\nu_\tau$) at a level of more than $15 \sigma$. Moreover, the KamLAND experiment has found that reactor $\ov{\nu}_e$'s disappear over a distance of about 180 km, also observing a distortion of their energy spectrum; while accelerator $\nu_\mu$'s  disappear over distances of 250 km (K2K experiment) and 735 km (MINOS experiment), again with a distortion in the energy spectrum.
All these results clearly imply that different flavored neutrinos are massive and mix according to what originally proposed several decades ago \cite{PMNS}.
Further possible evidence comes from the LSND experiment \cite{LSND}, reporting positive signals for $\ov{\nu}_\mu \rightarrow \ov{\nu}_e$ transitions, but such a result has not been confirmed by any other experiment.
Very recently, the MiniBooNE collaboration \cite{MINIBOONE} has reported the absence of  signal in the 475-3000 MeV energy range due to short-baseline ${\nu}_\mu \rightarrow {\nu}_e$ oscillations with a  squared mass difference $\Delta m^2$ compatible with the indication of $\ov{\nu}_\mu \rightarrow \ov{\nu}_e$ oscillations found in the LSND experiment.
Analogous missing events have been apparently observed in searches for $\ov{\nu}_\mu \rightarrow \ov{\nu}_e$ oscillations, although the MiniBooNE analysis of antineutrino data is not as precise as that of neutrino data because of a much lower statistics. Interestingly enough, the MiniBooNE collaboration has also reported an anomalous excess of low-energy $\nu_e$-like events, while a corresponding anomalous excess of antineutrino events has not been observed.

A possible explanation of the observed low-energy $\nu_e$ anomaly has been proposed \cite{GL08} in terms of very short-baseline disappearance of $\nu_e$'s due to oscillations into sterile neutrinos generated by a large squared mass difference (20 eV$^2\lesssim\Delta{m}^{2}\lesssim 330$ eV$^{2}$).
This hypothesis is motivated by the anomalous ratio between measured and predicted $^{71}$Ge production rates  observed in the Gallium radioactive source experiments GALLEX \cite{gallex} and SAGE \cite{sage}, and has been carefully confronted with the lack of any evidence of very short-baseline $\ov{\nu}_e$ disappearance in existing reactor experiments \cite{GL09}.
The thorough analysis on the experimental data reported in \cite{GL09} apparently implies a possible CPT violation in the neutrino sector.
Indeed, in very simple terms, the apparently observed $\nu_e$/$\ov{\nu}_e$ dissymmetry would point toward an asymmetry between the survival probability $P(\nu_e \rightarrow \nu_e)$ of electron neutrinos and that $P(\ov{\nu}_e \rightarrow \ov{\nu}_e)$ of electron antineutrinos against transitions into a sterile state, which is a clear signal of a CPT violation \cite{NuOsc}. At 99.71\,\% confidence level, quantity $A_{ee} \equiv P(\nu_e \rightarrow \nu_e) - P(\ov{\nu}_e \rightarrow \ov{\nu}_e)$
is negative, with a best-fit value of \cite{GL09}
\be \label{2}
A_{ee} = - \, 0.165 \, .
\ee
The possibility of a CPT violation in the neutrino sector is not a completely novel topic \cite{SME} \cite{KostMewes}, and experimental tests of such a violation have also been performed \cite{MINOSCPT} \cite{LSNDCPT}, with negative evidence. However, the result (\ref{2}) is very specific, and it is of some relevance to take it into due consideration because of its physical consequences. As a matter of fact, a CPT violation would signal, through the CPT theorem, a violation of the Lorentz symmetry, an issue that has been considered extensively in recent times \cite{LV}.
Let us also recall that neither Lorentz symmetry violation nor CPT violation automatically implies a non-vanishing asymmetry $A_{ee} \neq 0$, as also remarked in \cite{KostMewes} (while the contrary is always true, as abovesaid \cite{NuOsc}).
In the following we shall put forward a very simple CPT and Lorentz violating theoretical model for
neutrino oscillations keeping our phenomenological discussion as general as possible.

For the sake of simplicity, we consider only the mixing (with an angle $\theta$) between an active electron neutrino state $\nu_e$ and a sterile (under weak interaction) state $\nu_s$, with masses $m_1$ and $m_2$, respectively.
Standard simple calculations \cite{NuOsc} show that the survival probability for a beam of $\nu_s$ with momentum $p$ traveling for a time $t$ from the source can be generally cast in the following form:
\be \label{3}
P(\nu_e \rightarrow \nu_e) = 1 - \sin^2 2 \theta \, \sin^2 \frac{E_2 - E_1}{2} \, t \, .
\ee
In the usually considered ultrarelativistic case ($p \gg m$, distance $x \simeq t$), the energies $E_1$, $E_2$ of the two states approximate to
\be \label{4}
E_{1,2} \simeq p + \frac{m_{1,2}^2}{2 p} \, ,
\ee
so that we recover the standard formula
\be \label{5}
P(\nu_e \rightarrow \nu_e) \simeq 1 - \sin^2 2 \theta \, \sin^2 \frac{\Delta m^2}{4 p} \, x
\ee
with $\Delta m^2 \equiv m_2^2 - m_1^2$.
The same formula (\ref{5}) applies also for antineutrinos, $P(\nu_e \rightarrow \nu_e) = P(\ov{\nu}_e \rightarrow \ov{\nu}_e)$, since the change of sign in $E_1$, $E_2$ has no effect on the survival probability in (\ref{3}) or (\ref{5}).

At variance with the standard approach, let us assume that the dispersion relation between neutrino (and, possibly, also for other particles) energy and momentum is not that leading to Eqs. (\ref{4}), $E^2 = p^2 + m^2$, but is modified by effects beyond the Standard Model.
Such a possibility, indeed, has been extensively investigated by means of quite different approaches, sometimes extending, sometimes abandoning the formal and conceptual framework of Einstein's Special Relativity.
Moreover, some authors suspect that the Lorentz symmetry breaking may play a role in extreme astrophysical phenomena as, e.g., the observation of ultra-high energy cosmic rays with energies \cite{UHECR} beyond the Greisen-Zatsepin-Kuzmin \cite{GZK} cut-off, and of gamma rays bursts
with energies beyond 20 TeV originated in distant galactic sources \cite{Markarian}.
%\ c) longitudinal evolution of air showers produced by ultra-high energy hadronic
%particles which seem to suggest that pions are more stable than expected\cite{Showers}.
Lorentz-breaking observable effects appear \cite{LVN} in Grand-Unification Theories, in M-Theory and String theories, in (Loop) Quantum Gravity, in foam-like quantum spacetimes; in spacetimes endowed
with a nontrivial topology or with a discrete structure at the Planck length, or with a (canonical or noncanonical) noncommutative geometry; in so-called ``effective field theories'' and ``extensions'' of the Standard Model including Lorentz violating dimension-5 operators; in theories with a variable speed of light or variable physical constants. An interesting theoretical approach to Lorentz symmetry violation is found in DSR \cite{GAC,NCG,Deformed} working in $k$-deformed Lie-algebra noncommutative
($k$-Minkowski) spacetimes, in which both the Planck scale and the speed of light act as characteristic scales of a 6-parameter group of spacetime 4-rotations with deformed but preserved Lorentz symmetries.
In any of these theories, the most important consequence of a Lorentz violation is the modification of the ordinary energy-momentum dispersion relation, at energy scales usually assumed to be of the order of the Planck energy, by means of additional terms which vanish in the low-momentum limit. Such terms may produce nice physical effects, explaining for example baryogenesis \cite{EPL}, and several constraints (in the neutrino sector) on them have been obtained \cite{MPLA} from existing data coming from laboratory experiments.
Consequences of a nonstandard dispersion law on neutrino oscillations has been analyzed in \cite{ChenYang} and \cite{Coleman}; observable effects of relativistic non-covariance
on the flavor eigenstates have been investigated in \cite{Various}.
            %\footnote{By assuming a suitable Lorentz-violating dispersion relation
            %for the mass eigenstates of electron neutrinos, Carmona and Cortes\cite{CC}
            %succeed in explaining the so-called ``tritium beta-decay anomaly",
            %i.e., the anomalous excess of decay events near the endpoint of
            %the electron energy spectrum (where nonrelativistic few-eV
            %neutrinos are produced) which yields a characteristic ``tail'' in
            %the Kurie plot}.

In many of the theories above mentioned, the modified dispersion relation takes the general form
\be \label{6}
E^2 = p^2 + m^2 - \lambda \, E \, p^2 \,,
\ee
where $\lambda$ is a Lorentz violation parameter (with the dimensions of an inverse energy) satisfying the relation $\lambda E \ll 1$. In such approximation, deducing $E$ from Eq. (\ref{6}) we obtain (neglecting $\lambda^2$-order terms)
\be \label{7}
E = \pm \sqrt{p^2 + m^2} \, - \, \frac{\lambda \, p^2}{2} \, ,
\ee
where, as usual, the upper (lower) sign refers to particle (antiparticle) states. In the ultrarelativistic limit achieved in the neutrino oscillation experiments, we thus have
\be  \label{8}
E \simeq \pm \left( p + \frac{m^2}{2p} \right) - \, \frac{\lambda \, p^2}{2} \, .
\ee
As for the mass, the $\lambda$ parameter is, in general, particle-dependent and, as shown in \cite{EPL} on very general grounds, the requiring of CPT invariance would lead to a change of sign for $\lambda$ when we pass from particle to antiparticle states. For our purpose, instead, we take a {\it different} view, and assume that CPT does not hold anymore and $\lambda$ can have the same value for particles and antiparticles. Consequently, the energy difference between $\nu_e$ and $\nu_s$ in Eq. (\ref{3}) is given by
\be  \label{9}
E_2 - E_1 \simeq \frac{\Delta m^2}{2p} - \frac{p^2 \Delta \lambda}{2} \equiv \frac{\Delta M^2(p)}{2p} \, ,
\ee
while, for $\ov{\nu}_e$ and $\ov{\nu}_s$,
\be  \label{10}
\ov{E}_2 - \ov{E}_1 \simeq - \frac{\Delta m^2}{2p} - \frac{p^2 \Delta \lambda}{2} \equiv - \frac{\ov{\Delta M}^2(p)}{2p} \, .
\ee
For simplicity we have introduced the parameters
\be  \label{11}
\begin{array}{rcl}
\displaystyle \Delta M^2(p) & \equiv & \displaystyle \Delta m^2 - p^3 \Delta \lambda \\
&=& \displaystyle \Delta m^2 \left[ 1 - \left( \frac{p^2}{\Delta m^2} \right)p\,\Delta\lambda \right] \, , \\ & & \\
\displaystyle \ov{\Delta M}^2(p) & \equiv & \displaystyle \Delta m^2 + p^3 \Delta \lambda \\ &=& \displaystyle \Delta m^2 \left[ 1 + \left( \frac{p^2}{\Delta m^2} \right)p\,\Delta\lambda \right] \, ,
\end{array}
\ee
that can be interpreted as ``effective'', momentum-dependent, squared mass differences. Note that, since $p \gg m$, %and $\lambda \, p \ll 1$,
the ``correction'' terms in Eqs. (\ref{11}), that is $\displaystyle \left( \frac{p^2}{\Delta m^2} \right)p\,\Delta\lambda$, may {\it not} in principle be vanishingly small, depending on how much ultrarelativistic is the regime considered. Since $\ov{\Delta M}^2(p) \neq \Delta {M}^2(p)$, the survival probabilities for neutrino and antineutrino states
\begin{eqnarray}
P(\nu_e \rightarrow \nu_e) & \simeq & 1 - \sin^2 2 \theta \, \sin^2 \frac{\Delta M^2(p)}{4 p} \, x \, \label{12} \\
P(\ov{\nu}_e \rightarrow \ov{\nu}_e) & \simeq & 1 - \sin^2 2 \theta \, \sin^2 \frac{\ov{\Delta M}^2(p)}{4 p} \, x \, \label{13}
\end{eqnarray}
result to be {\it not} identical. Let us remark that the effect introduced here phenomenologically, although driven by a violation of the Lorentz symmetry, is substantially different from analogous effects introduced earlier by other authors \cite{Coleman}, with different phenomenological predictions (and in absence of a CPT violation), as a simple comparison shows.

Very roughly, in order to explain the apparently observed asymmetry (\ref{2}), the model proposed may be implemented as follows. For simplicity, let as assume that for ordinary $\nu_e$'s we have $\lambda_e \approx 0$, while for the sterile neutrino the $\lambda$ parameter is significatively different from zero: $\Delta \lambda \simeq - \lambda_s$. If the non-observation of an anomalous excess of low-energy antineutrino events, reported by the MiniBooNE collaboration, will be confirmed, it means that, in the given energy range,  $P(\ov{\nu}_e \rightarrow \ov{\nu}_e) \simeq 1$, that is $\ov{\Delta M}^2(p) \simeq 0$ or, from (\ref{11}),
\be \label{14}
\lambda_s \simeq \frac{\Delta m^2}{p^3} \, .
\ee
Because of Eq.\,\ref{11} this relation automatically implies $\Delta {M}^2 \simeq 2 \Delta {m}^2$, so that the survival probability for neutrinos, $P(\nu_e \rightarrow \nu_e)$, is significatively different from 1, thus explaining the observed anomalous excess of low-energy $\nu_e$-like events. Note that such a scenario is favored for ``low-energy'' events since, otherwise, the formula (\ref{14}) would require an exceedingly small $\lambda_s$.

Remarkably, if Eq.\,(\ref{14}) holds true, $P(\nu_e \rightarrow \nu_e)$ has a form and an effective energy dependence similar to the standard one in (\ref{5}), but with $\Delta m^2$ replaced by $2 \Delta m^2$. This allows a very rapid estimate of the order of magnitude for the $\lambda_s$ parameter. By using \cite{GL09} $\Delta m^2 \sim 100$ eV$^2$, $p \sim 1 $ GeV, from Eq. (\ref{14}) we obtain
\be  \label{15}
\lambda_s^{-1} \simeq 10^{16} \, \mbox{GeV} \, ,
\ee
which is a completely reasonable estimate once assumed that the Lorentz violation is induced by effects taking place, while approaching the Planck energy scale, around the Grand-Unification scale.

Summing up, we have proposed a very simple phenomenological model which predicts a CPT-violating asymmetry between particle and antiparticle states in neutrino oscillations involving a sterile neutrino. If the recent analysis in \cite{GL09} is further confirmed by novel experimental results, our model would imply that the standard energy-momentum dispersion relation for this sterile neutrino is modified according to Eq.\,(\ref{6}), with a CPT/Lorentz violation parameter $\lambda$ of the order of the Grand-Unification energy scale. This would directly affect the survival probabilities for neutrino and antineutrino oscillations (\ref{12}), (\ref{13}), with a possible extreme suppression of antineutrino transitions with respect to neutrino oscillations which, instead, would take place at a sizeable level.

Future neutrino oscillation experiments will allow to compare the present model with the standard scenario of flavor (or other) oscillations. Indeed, the dependence in our model of $P(\nu_e \rightarrow \nu_e)$ and $P(\ov{\nu}_e \rightarrow \ov{\nu}_e)$ on the energy of the neutrino beam is {\it different} from the standard case [compare Eqs. (\ref{12}), (\ref{13}) with Eq. (\ref{5})], and this may be easily tested in experiments with different baselines and energy ranges. We then expect that further interesting results may be achieved in a near future, along the lines depicted in the present paper.

%%%%%%%%%%%%%%%%%%%%%%%%%%%%%%%%%%%%%%%%%%%%%%%%%%%%%%%%%%%%%%%%%%%%%%%%%%%%%%%%%%%%%%%%%%%%%%%%%%%%%

\

\

\noindent {\bf Acknowledgements}

\noindent The authors are indebted with M. Laveder for his suggestions to work about the topic of the present paper and for very useful discussions.

%%%%%%%%%%%%%%%%%%%%%%%%%%%%%%%%%%%%%%%%%%%%%%%%%%%%%%%%%%%%%%%%%%%%%%%%%%%%%%%%%%%%%%%%%%%%%%%%%%%%%

\end{document}